\documentstyle[12pt]{article}
\def\be{\begin{equation}}
\def\ee{\end{equation}}
\def\ba{\begin{eqnarray}}
\def\ea{\end{eqnarray}}
\def\Xb{\bar{X}}
\def\pb{\bar{p}}
\def\a{\alpha}
\def\t{\theta}
\def\tb{\bar{\theta}}
\def\r{\rho}
\def\rb{\bar{\rho}}
\def\e{\epsilon}
\def\n00{N_{00}}
\def\n01{N_{01}}
\def\n11{N_{11}}
\begin{document}
\renewcommand{\theequation}{\thesection.\arabic{equation}}
\newcommand{\beq}{\begin{equation}}
\newcommand{\eeq}[1]{\label{#1}\end{equation}}
\newcommand{\ber}{\begin{eqnarray}}
\newcommand{\eer}[1]{\label{#1}\end{eqnarray}}
\begin{titlepage}
\begin{center}
         June, 1992           \hfill    IASSNS--HEP--92/33\\

\vskip .3in

{\large \bf On String Field Theory and Effective Actions}
\vskip .3in

{\bf Amit Giveon} \\ \vskip .1in

{\em School of Natural Sciences\\
     Institute for Advanced Study\\
     Princeton, NJ 08540}\footnotemark \\ \vskip .1in

\footnotetext{e-mail address: GIVEON@IASSNS.bitnet}
\end{center}

\vskip .2in

\begin{center} {\bf ABSTRACT } \end{center}
\begin{quotation}\noindent
A truncation of string field theory is compared with
the duality invariant effective action of
$D=4, N=4$ heterotic strings  to cubic order.
The  three string vertex must satisfy a set of  compatibility conditions.
Any cyclic three string vertex is compatible with the $D=4, N=4$
effective field theory.
The effective actions may be useful in understanding the non--polynomial
structure and the underlying symmetry
of covariant closed string field theory, and in addressing
issues of background independence.
We also discuss the effective action and string field
theory of the $N=2$ string.
\end{quotation}
\end{titlepage}
\vfill
\eject \noindent
\section{Introduction}
\setcounter{equation}{0}
The  low--energy effective action
of string theory is derived from string
field theory (SFT) by integrating out massive modes of the string field
(see for example ref.~\cite{GSW} for a review).
In SFT
different classical solutions correspond  to  string propagation  in
different backgrounds.
A property of string theory
is the possibility to deform a given vacuum continuously.
In four dimensions such deformations correspond to changing the
vacuum expectation values of massless scalars with a flat effective
potential.

Modes that are not massless in a given vacuum, may become light at some
other points of the moduli space of string backgrounds.
To obtain effective
theories compatible with the global structure of the moduli space of
backgrounds, one should integrate out only ultra--massive modes,
that is, modes that remain
massive in all backgrounds.
Alternatively, one may use restrictions deduced  from
space--time supersymmetry in order to find the
structure of the effective actions \cite{GP}.

A (successful) covariant string field theory of closed strings has been
discussed recently \cite{SFT,KS}.
Yet, the invariance principle that needs
a non--polynomial nature of the
covariant SFT
and the issue of background independence are not clear. Under the weak
assumptions of ref. \cite{SZ}, a covariant
string field theory of closed strings
is necessarily non--polynomial just as Einstein's theory of
gravity. The necessity of
higher--point interactions may be an indication of
the existence of a nontrivial invariance principle that needs a nonlinear
expression to become manifest.

Background independence means that  a SFT formulated
around a given background is equivalent to a SFT formulated around any
other backgrounds.
In other words, different classical solutions of a given
SFT should interpolate between string field theories formulated in
different backgrounds. Practically, however,
deforming a classical string field solution relating two backgrounds is
hard (see for example ref. \cite{Sen}).
This is mainly  due to the existence of an infinite number
of string oscillation modes.

As any off--shell formulation,
string field theory is not uniquely defined.
The three string interaction is described by the three string vertex, and
the vertex is defined by an off--shell formulation of the three--point
function in the $2-d$ conformal field theory.
A symmetric overlap three string vertex appears to be a canonical choice
that is simple, but it would be interesting to examine the other
possibilities. Even a symmetric vertex can be
defined by  different maps of three semi--infinite cylinders
(or unit discs) to the sphere.
(String field theories formulated  with different vertices appear to
be related  by a field redefinition~\cite{Eskin}.)\\

In this paper we study the compatibility conditions of string field
theory and effective actions of point particles to cubic order.
This can be used as a `test' for different formulations of SFT, and may be
helpful in understanding the non--polynomial structure in covariant
SFT of closed strings and its
(possible) invariance principle, as well
as  addressing  issues of background independence.

The effective actions
discussed  in this work  are  particular ones.
We start in section 2 with the $D=4$ heterotic string in toroidal
backgrounds. In four dimensions the effective action is a gauged $N=4$
supergravity with matter. The action is restricted as a consequence of
space--time supersymmetry,
and one is only left with the freedom to choose
the set of matter multiplets and the gauge algebra. Keeping a one to one
correspondence with the string states that may become light at some point
of the moduli space of toroidal background,
the number of matter multiplets
is infinite, and an infinite dimensional gauge symmetry
(called the `duality invariant string gauge algebra') is introduced
\cite{GP}.
A short review of a completely duality invariant effective action of $N=4$
heterotic strings is presented in section 2.

Once the structure constants of the gauge algebra are given,
the scalar potential is fixed. Expanding the scalar potential in physical
fields around zero
cosmological constant minima gives rise to a non--polynomial potential.
The quadratic term gives the correct mass spectrum of the corresponding
string modes; this is described in ref.~\cite{GP}.
In this work we focus on the
cubic terms and compare them
to the cubic interaction in string field theory.

In section 3 we present the three string vertex and the string field in
toroidal backgrounds. We then truncate the string field to modes that may
become light somewhere in the moduli space of toroidal backgrounds.
Compatibility with the effective action of $N=4$ heterotic strings is then
translated into a set of conditions,
that should be satisfied by the lower modes of
the three string vertex (namely, by the Neumann coefficients
$N_{00},N_{01}$ and $N_{11}$). In section 4 we prove that any cyclic
three string vertex obey the conditions, and therefore, a cyclic vertex
is compatible with the effective action.

In section 5 we discus the effective action of the $N=2$ string
in comparison with the $N=2$ string field theory.
In toroidal backgrounds, a  truncation of the $N=2$ string field
to modes that may become light  include, essentially, all
the modes that may become physical in toroidal backgrounds.
This is peculiar to $N=2$ strings which, therefore,
may be used as a simple
laboratory to more complicated string theories.

In section 6 we present a discussion  about this work
and some open questions. Finally, in a set of
appendices we present some useful formulas, and discuss two
examples.

\section{A Completely Duality Invariant Effective Action of $N=4$
Heterotic Strings}
\setcounter{equation}{0}
In this section we present a short review
of a completely duality invariant
effective action.
We then expand the scalar potential in physical fields to
cubic order.

The  construction of a completely duality invariant effective
action of the
$D=4$, $N=4$ heterotic string is presented in ref. \cite{GP}
using the structure of gauged $N=4$ supergravity with matter \cite{Wd},
obtained through superconformal methods by coupling a number of $N=4$
vector multiplets to the $N=4$ superconformal gauge multiplets.
Here we
present only some points relevant for the discussion in section 3; for
more details we refer the reader to ref.~\cite{GP}.

The
important point is that the form of a gauged $D=4$, $N=4$ supergravity
(with at most two space--time derivatives)
coupled to matter multiplets is
almost uniquely fixed:
it is completely determined by the knowledge of the
gauge group \cite{Wd}. Thus one has only the freedom to choose the
right set of matter multiplets and the correct gauge group in order to
figure out completely the low--energy effective action of the string.

The scalar fields which are relevant
for $D=4$ low--energy physics $Z_a^S$,
$a=1,...6$, $S=1,...,\infty$ will be labeled as follows:
\ba
Z_a^b &,& \; Z_a^I\;\; , \;\; Z_a^p \;\; ,\nonumber\\
p\in \Gamma^{6,22},\qquad  p^2=2 &,&
\qquad a,b=1,...,6, \qquad I=1,...,22,
\label{defz}
\ea
where $\Gamma^{6,22}$ is an even self--dual
lorentzian lattice of signature
(6,22). The momentum $p=(p_L,p_R)$ has 6 left--handed components $p_L^a$,
$a=1,...,6$, and 22 right--handed components $p_R^I$, $I=1,...,22$.
This is
the Narain lattice of the string toroidal compactification \cite{N}. The
scalar product is lorentzian
\be
pq=-\sum_{a=1}^6 p_L^a q_L^a + \sum_{I=1}^{22} p_R^I q_R^I,
\label{pq}
\ee
where L (R) denote left--handed (right--handed) momenta.
The indices $(b,I)$ refer to Cartan sub--algebra,
while the $p$ indices are
lorentzian length two generalized roots (and therefore their number is
infinite).

The structure constants of the duality invariant
gauge algebra which are relevant for $D=4$
low--energy physics are
\ba
f_{pqr}&=&\e(p,q)\delta_{p+q+r,0}, \qquad f_{Ipq}=p_I\delta_{p+q,0},
\qquad f_{apq}=p_a\delta_{p+q,0}, \nonumber \\
p^2&=&q^2=r^2=2.
\label{f}
\ea
The two cocycle $\e(p,q)$ satisfies the identities
\ba
\e(p,q)\e(p+q,r)=\e(p,q+r)\e(q,r), \nonumber\\
\e(p,q)=(-1)^{pq}\e(q,p), \qquad \e(p,-p)=1, \nonumber\\
\e(p,q)=\e(q,-p-q).
\label{e}
\ea

The structure constants $f_{RST}$ in (\ref{f}) are completely
antisymmetric.
Indices are  lowered and raised with the metric $\eta_{ST}$
defined by
\be
\eta_{ab}=-\delta_{ab}, \qquad \eta_{IJ}=\delta_{IJ}, \qquad
\eta_{pq}=\delta_{p+q,0}.
\label{eta}
\ee

In the Poincar\'e gauge it turns out that six of the vector multiplets
serve as ``compensating'' multiplets, absorbing redundant superconformal
symmetries.
The compensators $Z_a^b$ in (\ref{defz}) can be solved in terms
of the physical fields $Z^I_a$ and $Z^p_a$. This is done by using the
$SO(6)$ symmetry and the quadratic constraint
(originated by selecting the
Poincar\'e gauge and eliminating the
Lagrange multiplier auxiliary fields)
\be
\eta_{ST}Z_a^SZ_b^T=-\delta_{ab},
\label{quacon}
\ee
where $\eta_{ST}$ is given in (\ref{eta}).
Therefore, the fields $Z_a^S$ can
be regarded as the first six rows of an $SO(6,\infty)$ matrix.

The complete action of a gauged $N=4$ supergravity coupled to matter is
given in ref. \cite{Wd}. Here we present only the scalar potential
\be
V(Z)=\frac{1}{4} Z^{QU} Z^{SV}(\eta^{TW}+\frac{2}{3}Z^{TW})f_{QST}f_{UVW},
\label{V}
\ee
where
\be
Z^{QS}=\sum_a Z_a^Q Z_a^S.
\ee

The simplest zero cosmological constant minima of the potential~(\ref{V})
lie in the Cartan sub--algebra. Let us expand the scalar fields around a
vacuum expectation value (VEV) in the Cartan sub--algebra,
\begin{equation}
Z_a^S = C_a^b \delta_b^S + C_a^I \delta_I^S + z_a^S,
\label{13}
\end{equation}
where $C_a^b, C_a^I$, $a,b=1,..,6, \; I=1,..,22$, are constants and the
constraint~(\ref{quacon}) is satisfied.
A rotation of $Z_a^S$ to
\begin{equation}
(Z')_a^{S'} = Z^{S}_a M_{\;\;S}^{S'},
\label{14}
\end{equation}
where $M\in SO(6,\infty)$, preserves the scalar quadratic
constraint~(\ref{quacon}).
Changing the VEV of the scalar fields to new ones
in the Cartan sub--algebra  can be done by
a rotation  in $SO(6,22)\subset SO(6,\infty)$.

Under the orthogonal transformation~(\ref{14}),
the potential $V$  transforms into
\begin{eqnarray}
V' &=&  \frac{1}{4} (Z')^{QU} (Z')^{SV}
(\eta^{TW} + \frac{2}{3} (Z')^{TW}) f_{QST} f_{UVW} \nonumber \\
&=& \frac{1}{4} Z^{QU} Z^{SV}
(\eta^{TW} + \frac{2}{3} Z^{TW}) f'_{QST} f'_{UVW},
\label{15}
\end{eqnarray}
where
\begin{equation}
f'_{QST} = (M^{-1})^{\;\; T'}_T
(M^{-1})^{\;\; Q'}_Q f_{Q'S'T'} M^{S'}_{\;\; S}.
\label{16}
\end{equation}
Thus changing the VEV of the Higgs fields in the Cartan sub--algebra
is equivalent to a
transformation of the structure constants~(\ref{f}),
given by an $SO(6,22)$
rotation of $\Gamma^{6,22}$
(i.e. a rotation of the ``momenta'' labels $p$).
This transformation is an isomorphism of the gauge algebra.

Using the field redefinition  described above, one can always choose
a zero cosmological constant minimum to lie at
\begin{eqnarray}
Z^b_a&=&\delta_a^b,\;\;\;\; a,b=1,..,6,
\nonumber \\
Z^I_a&=&0,\;\;\;I=1,...,22,\;\;\; Z^p_a=0\;\;\forall p\in\Gamma^{6,22}.
\label{vev}
\end{eqnarray}

Expanding the scalar potential (\ref{V}) around the point (\ref{vev}) one
finds the quadratic and cubic terms
\ba
V(z)&=&
\frac{1}{2}\sum_{p,a,b}[(p^ap^a)z_b^p z_b^{-p}-p^ap^b z_a^p z_b^{-p}]
\nonumber\\&+&
\frac{1}{2} \sum_{p,q,r,a,b,c} \delta_{p+q+r,0}\;
\e(q,p)(p-r)^b \delta^{ac} z_a^p z_b^q
z_c^r \nonumber\\&+&
\sum_{p,I,a,b,c} p^I p^a \delta^{bc} z_c^{-p}(z_b^p z_a^I-z_a^p z_b^I) +
o(z^4).
\label{V(z)}
\ea

The duality invariant string gauge algebra requires the introduction of
extra fields, in order to complete the algebra defined by the structure
constants (\ref{f}) to a Lie algebra. These extra fields replace higher
spin fields of the string spectrum, and they are ultra--massive for all
backgrounds
(namely, for any VEV of the scalar fields). We, therefore, have
truncated  out these fields
by setting $z=0$ in $V(z)$ for fields that are
ultra--massive at all toroidal compactifications.

The quadratic term in (\ref{V(z)})
gives the correct mass spectrum of the  corresponding
string modes \cite{GP}. The field $p^az_a^p$ in $V(z)$ is the Goldstone
boson. In the next section we will check the compatibility
of the cubic terms in $V(z)$ (\ref{V(z)})
with the three string vertex in string field theory.

\section{The Three String Vertex and Effective Cubic Interaction}
\setcounter{equation}{0}
In this section
we present the three string vertex and the string field in
toroidal backgrounds. We then derive the effective cubic interaction for
$N=4$ heterotic strings, and study compatibility conditions
with the effective
action described in section 2. Finally, we discuss two examples.

\subsection{The Three String Vertex and the String Field in Toroidal
Backgrounds}
The three string vertex
in string field theory (SFT) is derived from the
off--shell definition of the three--point functions on the sphere.
We regard the
sphere as the complex plane with a point at $\infty$, and introduce a
coordinate $z$.
Any choice of three punctures are conformally equivalent to
one another, so one can take $z=0,1,\infty$ (for a cyclic vertex it is
sometimes more
convenient to choose different punctures;
this is discussed in section 4).
Correspondingly we introduce
three local coordinates $z_i(z)(i=1,2,3)$
as three local functions of $z$. We
arrange that each puncture corresponds
to the origin of a local coordinate, i.e.
\begin{equation}
z_1(0)=0,\qquad z_2(1)=0,\qquad z_3(\infty)=0.
\label{zzz}
\end{equation}
If the vertex is cyclic
(namely, invariant under a cyclic transformation of
1,2,3), the coordinates $z_2$ and $z_3$ are defined in terms of $z_1$ as
\cite{SZ}
\begin{equation}
z_2(z)=z_1(1-\frac{1}{z}),\qquad
z_3(z)=z_2(1-\frac{1}{z})=z_1(\frac{1}{1-z}).
\label{z123}
\end{equation}
We introduce a parameter $a$ given by the expansion of $z_1(z)$:
\begin{equation}
z_1(z)=az+...
\label{az}
\end{equation}

The off--shell three--point function is now defined to be
\begin{equation}
V_{ijk}=<0|h_3(V_i(0))h_2(V_j(0))h_1(V_k(0))|0>,
\label{Vijk}
\end{equation}
where $h_i$ are the inverse maps
\begin{equation}
h_i\equiv (z_i^{-1},\bar{z}_i^{-1}).
\label{h}
\end{equation}
When the vertex
operators   $V_i$ in (\ref{Vijk}) correspond to primary fields,
the off--shell three--point function is
\begin{equation}
V_{ijk}={\rm lim}_{\epsilon\rightarrow 0}\;
h'_3(\epsilon)^{d_i}h'_2(\epsilon)^{d_j}
h'_1(\epsilon)^{d_k}<0|V_i(h_3(\epsilon))V_j(h_2(\epsilon))
V_k(h_1(\epsilon))|0>,
\label{vijk}
\end{equation}
where $h'=|\partial h|^2$, and $d_i$ is the
conformal weight of the vertex $V_i$.

The cubic interaction in string field theory is given by
\be
S_3=\frac{g}{3}\;{}_{123}<V||\Psi>_1|\Psi>_2|\Psi>_3,
\label{S3}
\ee
where $|\Psi>$ is the string field,
and ${}_{123}<V|$ is the three string
vertex. The vertex is constructed such that
when contracted with three string
field modes it gives the corresponding off--shell three--point function
$V_{ijk}$ in (\ref{Vijk}).
In toroidal background,
the three string vertex can be given in the form
(see for example \cite{KZ}):
\be
{}_{123}<V|={}_{123}<0|e^{E_{123}}\epsilon(p_2,p_3)(...).
\label{V123}
\ee
The expression for $E_{123}$ is:
\be
E_{123}=\frac{1}{2}\sum_{r,s=1,2,3}\sum_{n,m\geq 0}
(\sum_{a} N_{nm}^{rs} \a_n^{a(r)}\a_m^{a(s)} +
 \sum_{I} {\bar N}_{nm}^{rs} \bar{\a}_n^{I(r)}\bar{\a}_m^{I(s)}).
\label{E}
\ee
The index $a$ refers to left--moving coordinates
and the index $I$ refers to
right--moving coordinates similar to the notation used
for the effective action
in section 2. The operators $\a^{a(r)}_n$ in (\ref{E}) are
the modes of the left--handed coordinate $X_L^a$ of the $r$'th string;
the operators $\bar{\a}^{I(r)}_n$ are the modes of the
right--handed coordinate $X_R^I$ of
the $r$'th string. The modes obey the commutation relations
\be
[\a_n^{a(r)},\a_m^{b(s)}]=n\eta^{ab}\delta^{rs}\delta_{n+m,0},\qquad
[\bar{\a}_n^{I(r)},\bar{\a}_m^{J(s)}]
=n\eta^{IJ}\delta^{rs}\delta_{n+m,0}
\label{aa}
\ee
The zero mode operators get values in the even--self dual lorentzian
lattice when acting on $|0>_{123}$ (see appendix A), namely,
\be
(\a_0^{a(r)}(p_r),\bar{\a}_0^{I(r)}(p_r))\equiv(p_L^{a(r)},p_R^{I(r)}),
\label{a=p}
\ee
where $(p_L,p_R)$ is a lorentzian momentum in the Narain lattice of
compactification.
The two cocycle $\epsilon(p_r,p_s)$ in (\ref{V123})
obeys the relations (\ref{e}); the
momenta $p_r$ are the lorentzian momenta of the $r$'th string.
Finally, the dots in (\ref{V123}) refer to
delta functions of left--right
level matching conditions ($L_0=\bar{L}_0$), and momentum conservation
($p_1+p_2+p_3=0$), and to ghost factors which are not important for our
discussion.

The Neumann coefficients $N_{nm}^{rs}$
(and their  complex conjugates ${\bar N_{nm}^{rs}}$)
in $E_{123}$ (\ref{E}) depend on the off--shell
definition of the three point function (\ref{Vijk}), namely, they are
defined in terms of the maps $h_i$ (\ref{h}). The
Neumann coefficients $N_{nm}^{rs}$ with $n,m >0$ are
invariant under $SL(2,C)$ transformations of $z$ \cite{LPP}.
The coefficients $N_{00}^{rs}$, $N_{0m}^{rs}$, and $N_{n0}^{rs}$
can be redefined
by making use of momentum conservation.
With the use of momentum conservation
these coefficients are also $SL(2,C)$ invariant. The  formulas for the
Neumann coefficients of the closed string vertex
(as contour integrals over functions of $h$ and $h'$)
appear for example in \cite{KS}.
For a cyclic vertex, the $N$'s are given
in terms of one function $h$ (see section 4).

In toroidal backgrounds,
the string field $|\Psi>$ can be expanded in terms
of the Fock space states
\be
|\Psi>=|T>+|Z_L>+|Z_R>+|Z_{LR}>+...
\label{P}
\ee
(Here we ignore the ghosts in $|\Psi>$.)
The state $|T>$ corresponds to the tachyon mode
\be
|T>=\sum_{p^2=0}T_p|p>,
\label{T}
\ee
where $p^2=p_R^2-p_L^2$.
(The tachyon does not appear when we discuss the
heterotic string).
The states $|Z_L>$ correspond to the first left--moving excitations
\be
|Z_L>=\sum_{a,p^2=2}Z_p^a{\a}_{-1}^a|p>,
\label{ZL}
\ee
and the states $|Z_R>$ correspond to the first right--moving excitations
\be
|Z_R>=\sum_{I,p^2=-2}Z_p^I{\bar \a}_{-1}^I|p>.
\label{ZR}
\ee
(The states $|Z_R>$ do not appear when we discuss the heterotic string).
Similarly, the states $|Z_{LR}>$ correspond to
\be
|Z_{LR}>=\sum_{a,I,p^2=0}Z_p^{aI}\a_{-1}^a {\bar \a}_{-1}^I|p>.
\label{ZLR}
\ee

In the Fock space expansions (\ref{T})--(\ref{ZLR})
we have restricted to states satisfying the
left--right level matching condition, and for that purpose we sum over
momenta with particular lorentzian length appropriate to each mode.
The modes which do not satisfy the
left--right level matching condition are projected out
anyway when contracted with $|V>_{123}$.
The dots in (\ref{P})
correspond to higher excitation modes of the string;
we did not keep them as they do not appear in
low--energy effective actions.
In other words, we truncate massive modes that remain massive at all
toroidal backgrounds.

To end this sub--section, we remark that a
truncation of massive modes by setting them to zero is consistent
if we deal with  cubic interactions only.
However, in case one is interested in
higher order interactions of light modes, the
propagation of ultra--massive modes usually contributes to the
amplitudes.
One should then integrate out the
ultra--massive modes properly in order to
derive the correct effective action.

\subsection{The Effective Cubic Interaction and Compatibility Conditions
for $N=4$ Heterotic Strings}
We are now ready to present the part of the cubic interaction (\ref{S3})
which is relevant for low--energy effective actions of $N=4$ heterotic
strings.
After some
straightforward calculations one finds
\ba
S_3&=&\frac{g}{3}\{{}_{123}<V||Z_L>_1|Z_L>_2|Z_L>_3 +
{}_{123}<V||Z_L>_1|Z_L>_2|Z_{LR}>_3 \nonumber\\
&+&{}_{123}<V||Z_L>_1|Z_{LR}>_2|Z_L>_3
+{}_{123}<V||Z_{LR}>_1|Z_L>_2|Z_L>_3)\}+...\nonumber\\
&=& \frac{g}{3}
\sum_{p_1,p_2,p_3,a,b,c}
\delta_{p_1+p_2+p_3,0}\; e^{E_{00}}\epsilon(p_2,p_3)
\Gamma_{abc} \{Z^a_{p_1}Z^b_{p_2}Z^c_{p_3} \nonumber\\
&+&\sum_{I,r=1,2,3}
({\bar N}_{10}^{1r}p_r^I Z^{aI}_{p_1}Z^b_{p_2}Z^c_{p_3}+
 {\bar N}_{10}^{2r}p_r^I Z^a_{p_1}Z^{bI}_{p_2}Z^c_{p_3}+
 {\bar N}_{10}^{3r}p_r^I Z^a_{p_1}Z^{b}_{p_2}Z^{cI}_{p_3})\},
\nonumber\\{}
\label{ZZZV}
\ea
where the lorentzian length is $p_r^2=2$ in $Z_{p_r}^a$ and $p_r^2=0$ in
$Z_{p_r}^{aI}$, and
\ba
\Gamma_{abc}=\sum_{r}(N_{11}^{31}N_{10}^{2r}p^b_r\eta^{ac}+
                      N_{11}^{32}N_{10}^{1r}p^a_r\eta^{bc}+
                      N_{11}^{21}N_{10}^{3r}p^c_r\eta^{ab})
\nonumber\\ +\beta
\sum_{r,s,t}N_{10}^{1r}N_{10}^{1s}N_{10}^{1t}p^a_r p^b_s p^c_t .
\label{Gabc}
\ea
In (\ref{Gabc}) $\beta=1\;(0)$ for the bosonic (heterotic) string;
this is explained later.
Some formulas that are useful for the calculation of (\ref{ZZZV}) are
given in appendix~A.

The factor $e^{E_{00}}$ in (\ref{ZZZV}) is given by
\be
E_{00}=\frac{1}{2}\sum_{r,s}(N_{00}^{rs} p_{L(r)}p_{L(s)}+
{\bar N}_{00}^{rs} p_{R(r)}p_{R(s)}).
\label{E00}
\ee
For a cyclic vertex one can use momentum conservation to bring
$N_{00}^{rs}$ into the form (see section 4):
\be
N_{00}^{rs}\equiv \delta^{rs}N_{00}=-\delta^{rs}{\rm log}a,
\label{N0}
\ee
where $a$ is defined in (\ref{az}). One finds
\be
e^{E_{00}}=a^{-\frac{1}{2}\sum_{r=1}^{3}p_{L(r)}^2}
    \; \bar{a}^{-\frac{1}{2}\sum_{r=1}^{3}p_{R(r)}^2}.
\label{eEE}
\ee
This `scaling--factor'
is a result of the $h'$ appearing in the off--shell
three--point function~(\ref{vijk})
(we call it a scaling--factor because
$a\rightarrow \lambda a$ under the scale transformation
$h\rightarrow \lambda^{-1} h$).\\

Next we discuss briefly the property of
a heterotic string field $|\Psi>$ (in the Neveu--Schwarz sector).
The string field in (\ref{P}) is presented for bosonic modes. For the
heterotic string, however,
the right--handed modes are  bosonic while
the left--handed modes correspond to a fermionic string.
Therefore, the expansion of the string field
in Fock space states includes
fermionic modes as well. The Fock space of a fermionic
string includes modes in different pictures (the picture presented in
(\ref{P}) is the 0--picture).
In addition, the string vertex includes both
bosonic and fermionic ghost factors. A proper cancelation of the ghost
factors requires that a three--point function is computed with two modes
in the ($-1$)--picture, and one mode in the 0--picture.
One then finds that
the only difference in the off--shell three--point function
compared  to a computation in the 0--picture is  the
cubic $ppp$ term in (\ref{Gabc}), namely:
$\beta=1$ for the bosonic  string while
$\beta=0$ for the heterotic string.

To compare with the duality invariant effective action of $N=4$ heterotic
strings the following should be done. The left--moving index $a$ is split
into an internal index $a=1,...,6$ and a space--time index $\mu=1,...,4$.
Similarly, the
right--moving index $I$ is split into the
internal index $I=1,...,22$ and a
space--time index. The $D=4$ space--time can be regarded as a
decompactification  limit
of the toroidal background. The space--time dependence of the fields
and the space--time derivatives then appear correctly after performing a
Fourier transform from momentum space to coordinate space. Here we
restrict the discussion to the scalar fields
$Z$ with internal indices only. We are thus comparing
the part of $S_3$ in (\ref{ZZZV}) that is
relevant for low--energy with the scalar potential $V(Z)$ (\ref{V}).
Moreover, we should restrict to fields $Z_p^{aI}$ with $p=0$ only, since
the fields $Z_p^{aI}$ with $p\neq 0$
remain ultra--massive in $D=4$ (they may become light
only in the decompactification limit of internal coordinates).

After doing all that, we find that manifest
compatibility with the scalar potential
(to cubic order) is translated into the following conditions for the
Neumann coefficients (modulo redefinitions of $N_{00}^{rs}$ and
$N_{01}^{rs}$ using momentum conservation; this will be
discussed in more  detail in section 4):
\be
N_{11}^{12}N_{01}^{33} + {\rm cyclic}=0,
\label{con1}
\ee
\be
N_{11}^{12}(N_{01}^{13}+N_{01}^{23})+{\rm cyclic}=0,
\label{con2}
\ee
\ba
({\bar N}_{01}^{13}-{\bar N}_{01}^{23})
{[}N_{11}^{12}(N_{01}^{13}-N_{01}^{23})+
   N_{11}^{23}(N_{01}^{11}-N_{01}^{21})&+&
   N_{11}^{13}(N_{01}^{12}-N_{01}^{22}){]} \nonumber\\
+\; {\rm cyclic}&=&0,
\label{con3}
\ea
\ba
N_{11}^{12}(N_{01}^{23}-N_{01}^{13})
\left( e^{\bar{N}_{00}}+ \bar{N}_{01}^{23} -
            \bar{N}_{01}^{13} \right) + {\rm cyclic} =0,
\label{con4}
\ea
\be
|a|=1 \qquad ({\rm i.e.}\quad {\rm Re}\;N_{00}=0),
\label{con5}
\ee
where $X(1,2,3)+{\rm cyclic}\equiv X(1,2,3)+X(2,3,1)+X(3,1,2)$ (this is
only done for the upper indices).

The first condition (\ref{con1}) means that a term
$p^b_2  \delta^{ac} z_a^{p_1} z_b^{p_2} z_c^{p_3}$
is absent in $V(z)$ (\ref{V(z)}) (the momenta $p,q,r$
in  (\ref{V(z)}) are replaced with $p_1,p_2,p_3$, respectively).
Such a term can always be removed by the use of momentum conservation
$p_2=-p_1-p_3$. In other words,
the  Neumann coefficients $N_{01}^{rr}$
can be redefined to $0$ by making
use of momentum conservation $p_1=-p_2-p_3$ for  the term $p_1\a^{(1)}$
in $E_{123}$ etc.,
and therefore,  condition (\ref{con1}) is always satisfied.

Conditions (\ref{con2})--(\ref{con4}) specify
the correct relative factors
between different cubic terms in $V(z)$.
The second condition  (\ref{con2}) means
that the relative factor between
the terms $p_1^b  \delta^{ac} z_a^{p_1} z_b^{p_2} z_c^{p_3}$ and
$p_3^b \delta^{ac} z_a^{p_1} z_b^{p_2} z_c^{p_3}$ in $V(z)$
is correct. However, this relative factor can always be fixed by using
momentum conservation
and the properties (\ref{e}) of the two cocycle which
lead to the identity
\be
\sum_{p_1,p_2,p_3,a,b,c} \delta_{p_1+p_2+p_3,0} \; \e(p_2,p_1)
(p_1+p_3)^b \delta^{ac}
z_a^{p_1} z_b^{p_2} z_c^{p_3} = 0.
\label{identity}
\ee

A three string vertex should therefore satisfy only the conditions
(\ref{con3})--(\ref{con5}).
The third
condition  (\ref{con3}) means that the relative factor between the terms
$z_a^p z_b^I$ and $z_b^p z_a^I$ in (\ref{V(z)}) is correct,
and the fourth
condition (\ref{con4}) means that the relative factor between the terms
$z_a^{p_1} z_b^{p_2} z_c^{p_3}$
and $  z_c^{-p} z_a^{p} z_b^{I}$ in $V(z)$
is correct.   (In deriving
condition (\ref{con4}) we have used the identity (\ref{identity})).

Conditions  (\ref{con3}),(\ref{con4}) are satisfied by any
cyclic three string vertex (this is shown in section 4).
The last condition, however,  is somewhat special:
cyclic vertices may violate only the condition
$|a|=1$ (\ref{con5}).
In other words, after solving $p_{R(r)}^2$ in terms
of $p_{L(r)}^2$,
the term (\ref{eEE}) gives rise to a factor
\be
|a|^{-\sum_{r=1}^3 p_{L(r)}^2},
\label{eE}
\ee
which is, apparently, the only
difference between effective cubic interactions
derived from different cyclic
three string vertices.~\footnote
{
When $p_L$ includes space--time momenta in addition to internal momenta,
then the factor (\ref{eE}) is 1 on--shell.
}
However, this factor can be eliminated
by performing an appropriate
field redefinition;~\footnote
{
This was pointed out to me by B. Zwiebach.
}
this is demonstrated within the examples below.

\subsection{Examples}
In this sub--section we discuss two examples.
The first is the Witten vertex \cite{W};
it corresponds to the choice (see for example ref. \cite{SZ})
\ba
z_1(z) &=&
i\alpha\frac{(1+t)^{3/2}-(1-t)^{3/2}}{(1+t)^{3/2}+(1-t)^{3/2}}=
\alpha\frac{3\sqrt{3}}{4}z+...,\nonumber \\
t &=& i\sqrt{3}\frac{z}{z-2}.
\label{witten}
\ea
This is the natural extension of the covariant open string field theory
vertex.
With the standard normalization $\alpha=1$, the entire sphere is
covered by
three unit discs $\{|z_i|\leq 1, i=1,2,3\}$, where $z_i$ are given in
(\ref{zzz}),(\ref{z123}).
With this normalization $a=\frac{3\sqrt{3}}{4}$
in (\ref{az}).

The Neumann coefficients for Witten vertex were computed for the open
string in \cite{GJ},
and can be computed for the closed string following
the discussion  in section 4 below (see appendix B).
In both cases the Neumann coefficients are
\be
N_{00}^{rs}=\delta^{rs}{\rm log}(\frac{4}{3\sqrt{3}}),
\label{n00}
\ee
\be
N_{01}^{21}=N_{01}^{32}=N_{01}^{13}=-N_{01}^{12}
=-N_{01}^{23}=-N_{01}^{31}
=\frac{2}{3\sqrt{3}},
\qquad N_{01}^{rr}=0,
\label{n01}
\ee
\be
N_{11}^{12}=N_{11}^{13}=N_{11}^{23}=\frac{16}{27},
\qquad N_{11}^{rr}=-\frac{5}{27}.
\label{n11}
\ee
The Neumann coefficients (\ref{n00}),(\ref{n01}),(\ref{n11})
obey the conditions
(\ref{con1})--(\ref{con4}). However, condition (\ref{con5}) is not
satisfied because $|a|\neq 1$ in (\ref{witten}).

Inserting the Neumann coefficients in $S_3$ (\ref{ZZZV}),
and truncating the string field to  the
modes $Z_a^p, Z_{p=0}^{aI}\equiv Z_a^I$,
one finds that
the cubic interaction in  a SFT based on Witten vertex  is
\ba
S_3(Z)=
\frac{g'}{3}\{\frac{1}{2}
\sum_{p_1,p_2,p_3,a,b,c} |a|^{-\sum_{r=1}^3 p_{L(r)}^2} \;
\times \nonumber\\
\delta_{p_1+p_2+p_3,0}\; \e(p_2,p_1)(p_1-p_3)^b \delta^{ac}
Z_a^{p_1} Z_b^{p_2} Z_c^{p_3}+ \nonumber\\
\sum_{p,I,a,b,c} |a|^{-2p_{L}^2}
p^I p^a \delta^{bc} Z_c^{-p}(Z_b^p Z_a^I-Z_a^p Z_b^I)\}.
\label{S(Z)}
\ea
Here we have used the lorentzian
scalar product (\ref{pq}) and the values $p^2=2$ in
$Z_a^p$ in order to solve $p_R^2$ in terms of $p_L^2$, and consequently
$g'=\frac{8^3}{9^4}g$.
Except for the factor $|a|^{-\sum_{r=1}^3 p_{L(r)}^2}$
(in the second sum
in (\ref{S(Z)}) $|a|^{-\sum_{r=1}^3 p_{L(r)}^2}=|a|^{-2p_{L}^2}$),
$S_3(Z)$ (\ref{S(Z)}) equals to the cubic term in
the scalar potential $V(z)$ (\ref{V(z)}).

The extra factor $|a|^{-\sum_{r=1}^3 p_{L(r)}^2}$
can be eliminated by an
appropriate field redefinition. This is done as follows: the truncated
action derived
from SFT is of the form
\ba
S(Z)&=&\frac{1}{2}\sum_p p^2 Z_p Z_{-p}\nonumber\\
&+& \frac{g}{3}\sum_{p,q} f(p,q)
e^{-A(p^2+q^2+(p+q)^2)} Z_p Z_q Z_{-p-q}.
\label{SSS}
\ea
By defining
\ba
Z_p&=&z_p-\frac{g}{3}\sum_{q}
\frac{{[}f(p,q) e^{-A(p^2+q^2+(p+q)^2)}-1{]}}{p^2} z_q z_{-p-q} \;\;
\;\;{\rm for}\;\; p\neq 0, \nonumber\\
Z_0&=&z_0,
\label{def}
\ea
the action (\ref{SSS}) becomes
\be
S(z)= \frac{1}{2}\sum_p p^2 z_p z_{-p}
+ \frac{g}{3}\sum_{p,q} f(p,q) z_p z_q z_{-p-q} + o(z^4).
\label{Sdef}
\ee
After performing this field redefinition, the action  derived from SFT
(\ref{S(Z)}), equals to the scalar potential $V(z)$ in
(\ref{V(z)}) to cubic order.

Nevertheless, the simplest way to get rid of the factor
$|a|^{-\sum_{r=1}^3 p_{L(r)}^2}$ is to define the three string vertex
with a map $z_1(z)=e^{i\phi}z+....$, namely, $|a|=1$.
A simple vertex of this type is the SCSV vertex~\cite{CSV};
it is defined by the map
\be
z_1(z)=z.
\ee
The map $z_3(z)=h_2(z)=\frac{1}{1-z}$ in (\ref{z123}),(\ref{h})
is the $SL(2,C)$
transformation sending $\infty\rightarrow 0,\; 0\rightarrow 1,\;
1\rightarrow \infty$. This vertex is cyclic but not symmetric under
all permutations.
The Neumann coefficients for the open string
were computed in \cite{L}, and
can be computed for the closed string
following the discussion in section 4
below (see appendix C). For the closed string the Neumann
coefficients are
\ba
N_{01}^{rr}&=&N_{11}^{rr}=0,\qquad
N_{00}^{rs} = \delta^{rs} {\rm log} i,\nonumber\\
N_{01}^{21}&=&N_{01}^{32}=N_{01}^{13}=i, \qquad
N_{01}^{12}=N_{01}^{23}=N_{01}^{31}=0,
\nonumber\\
N_{11}^{12}&=&N_{11}^{13}=N_{11}^{23}=-1.
\label{NCSV}
\ea
It turns out that only
conditions (\ref{con1}) and (\ref{con2}) are violated;
but these are not important, as explained in sub--section 3.2.

The SCSV vertex is an example of a cyclic
three string vertex that
satisfies conditions (\ref{con3})--(\ref{con5}).
Therefore, it is  compatible with the $N=4$ supergravity
coupled to matter
effective field theory.  In the next section
we show that all the cyclic vertices
are compatible with
the $N=4$ effective action.

\section{A Cyclic Vertex  is Compatible with Effective
Actions}
\setcounter{equation}{0}

In this section we prove that any cyclic three string vertex obey
the conditions (\ref{con3}),(\ref{con4}). Therefore, a cyclic vertex
is compatible with the effective action.

We start with a symmetric vertex, namely,
a vertex that is invariant under any permutation of the three strings.
In particular, the vertex is cyclic. To derive the conditions on the
functions $h_i$ in (\ref{h}), it would be simpler to work with the three
punctures at $z=1,\r,\rb$, where $\r=e^{\frac{2\pi i}{3}}$.\footnote
{
This was suggested to me by K. Ranganathan and B. Zwiebach.
}
In this case, a
cyclic permutation $P_{123}$
of  the punctures is given by a multiplication by
$\r$, and therefore,
$h_1,h_2,h_3$ are given in terms of a single function $h$:
\be
z=h_1(z_1)\equiv h(z_1),\qquad z=h_2(z_2)=\r h(z_2),
\qquad z=h_3(z_3)=\rb h(z_3),
\label{h123}
\ee
such that
\be
h_1(0)=1, \qquad h_2(0)=\r,\qquad h_3(0)=\rb .
\label{h(0)}
\ee

It is convenient to
define a function $O(z)$ and expansion parameters $a_1,a_2$ by
\be
h(z)=e^{O(z)},
\label{h=eo}
\ee
where
\ba
O(z)&=& a_1 z + a_2 z^2 + o(z^3),  \nonumber\\
h(z)&=& 1 + a_1 z + \frac{1}{2}a_1^2 z^2 + a_2 z^2 + o(z^3).
\label{a1}
\ea

The group  of permutations ${\bf S}_3$ is generated by the cyclic
transformation  $P_{123}$ and one permutation, say $P_{23}$.
The map $P_{23}$
exchanges $\r$ with $\rb$ leaving the point $z=1$ fixed. This map is
\be
P_{23}:\;\; z\rightarrow z'=\frac{1}{z}.
\label{z'}
\ee
The choice of $h_1$ cannot be arbitrary;
a constraint comes from $P_{23}$.
In terms of $z_1$ the map $P_{23}$ must be a phase rotation. Since
$P_{23}^2=1$, the only choice of phase is by a minus sign,
and therefore $h$ satisfies
\be
h(z)=\frac{1}{h(-z)}.
\label{hcon}
\ee
The condition (\ref{hcon}) implies that $O(z)$ is an odd function,
and in particular $a_2=0$.

We are now ready to calculate the Neumann coefficients. For the closed
string vertex, $N_{00}^{rs}$,
$N_{01}^{rs}$, $N_{11}^{rs}$ are given by (see for example \cite{KS})
\ba
N_{00}^{rr}&=&{\rm log} h'_r(0),\nonumber\\
N_{00}^{rs}&=&{\rm log} (h_r(0)-h_s(0)), \qquad {\rm for} \;\; r\neq s,
\nonumber\\
N_{01}^{rs}&=&N_{10}^{sr}=\oint \frac{dz}{2\pi i} \frac{1}{z} h'_s(z)
\frac{i}{h_r(0)-h_s(z)}, \nonumber \\
N_{11}^{rs}&=&\oint \frac{dz}{2\pi i} \frac{1}{z} h'_r(z)
             \oint \frac{dw}{2\pi i} \frac{1}{w} h'_s(w)
\frac{-1}{(h_r(z)-h_s(w))^2},
\label{Nint}
\ea
where each contour surrounds the origin of the integration
variable~\footnote{
There is a factor of $-i$ in $N_{01}$ and a factor of $-1$ in $N_{11}$
relative to ref. \cite{KS}; it comes from a relative $i$ factor in the
definition of $\a_{1}$ and $\a_0$.
}.

Using (\ref{h123}),(\ref{h(0)}) and (\ref{a1}) one finds for a symmetric
vertex
\ba
N_{00}^{11}&=&{\rm log}a_1,\qquad
N_{00}^{22}={\rm log}a_1 +\frac{2\pi i}{3}, \qquad
N_{00}^{33}={\rm log}a_1 -\frac{2\pi i}{3}, \nonumber \\
N_{00}^{21}&=&N_{00}^{12}+\pi i
= {\rm log} \sqrt{3}+\frac{5\pi i}{6}, \qquad
N_{00}^{23}=N_{00}^{32}+\pi i = {\rm log} \sqrt{3}+\frac{\pi i}{2},
\nonumber \\
N_{00}^{13}&=&N_{00}^{31}+\pi i = {\rm log} \sqrt{3}+\frac{\pi i}{6},
\label{N00}
\ea
\be
N_{01}^{rr}=-i\frac{1}{2}a_1,
\label{N01rr}
\ee
\ba
N_{01}^{21}=N_{01}^{13}=N_{01}^{32}
&=&\frac{a_1}{\sqrt{3}}e^{-\frac{\pi i}{3}},
\nonumber\\
N_{01}^{12}=N_{01}^{31}=N_{01}^{23}&=&
\frac{a_1}{\sqrt{3}}e^{-\frac{2\pi i}{3}},
\label{N01rs}
\ea
\be
N_{11}^{rs}=\frac{a_1^2}{3}   \qquad {\rm for}\;\; r\neq s
\label{N11}
\ee
In (\ref{N00})--(\ref{N11}) we have used the numerical values
$1-\r=|1-\r|e^{-\pi i/6}$ and $|1-\r|=\sqrt{3}$.
The parameter $a_1$ in (\ref{N00})--(\ref{N11}) is defined in eq.
(\ref{a1}). In deriving the result (\ref{N01rr}) we have used the fact
that $a_2=0$ in (\ref{a1});
the other results do not depend on the invariance of the vertex under
$P_{23}$ (i.e., cyclic symmetry
is sufficient to derive all but eq. (\ref{N01rr});
we will discuss the case $a_2\neq 0$ later).

To compare with the Neumann coefficients derived for Witten
vertex given in (\ref{n00}), (\ref{n01}), and
(\ref{n11}), we have to redefine the $N's$ in
(\ref{N00})--(\ref{N11}) by using $p_1+p_2+p_3=0$. (We know that the
Neumann coefficients must be invariant under $SL(2,C)$ transformations,
up to
a redefinition of $N_{00}^{rs}$ and $N_{01}^{rs}$
by making use of momentum
conservation \cite{LPP}). For $N_{00}^{rs}$ one finds
\ba
E_{00} &=& \frac{1}{2} \sum_{r,s=1,2,3}
(N_{00}^{rs} p_{L(r)} p_{L(s)} + \bar{N}_{00}^{rs} p_{R(r)}p_{R(s)})
\nonumber\\
&=& \frac{1}{2} {[}
({\rm log} \frac{a_1}{\sqrt{3}})\sum_{r=1}^3 p_{L(r)}^2
+ ({\rm log} \frac{\bar{a}_1}{\sqrt{3}})\sum_{r=1}^3 p_{R(r)}^2 {]},
\ea
and therefore, the redefined $N_{00}$ are
\ba
{\rm redefined}\;\; N_{00}&:&\nonumber\\
N_{00}^{rs}\equiv \delta^{rs}N_{00}
&=&\delta^{rs} {\rm log} \frac{a_1}{\sqrt{3}}.
\label{N00eff}
\ea
We have thus recovered eq. (\ref{N0}) if we  identify
\be
a=\frac{\sqrt{3}}{a_1}.
\label{aa1}
\ee
(The parameter $a$ is defined by expending $z_1=h_1^{-1}(z)$ (\ref{az}),
and therefore is
proportional to the inverse of $a_1$. An illustration of the relation
between the parameters $a$ and $a_1$ is given for Witten  vertex
in appendix B.)

In the same way,
by using $p_1+p_2=-p_3,\;p_1+p_3=-p_2,\;p_2+p_3=-p_1$ we
can redefine $N_{01}^{rs}$ s.t. $N_{01}^{rr}=0$.
One finds
\be
\sum_{r,s=1,2,3} N_{01}^{rs} p_r \a^{(s)}_1=
\frac{1}{2a}\{ \a^{(1)}_1 (p_2-p_3)+\a^{(2)}_1 (p_3-p_1)
+\a^{(3)}_1(p_1-p_2)\},
\ee
and therefore, the redefined $N_{01}$ are
\ba
{\rm redefined} \;\;N_{01}&:& \nonumber \\
N_{01}^{21}=N_{01}^{13}=N_{01}^{32}
&=&-N_{01}^{12}=-N_{01}^{31}=-N_{01}^{23}=
\frac{a_1}{2\sqrt{3}}=\frac{1}{2}a^{-1}=\frac{1}{2}e^{N_{00}},
\nonumber \\
N_{01}^{rr}&=&0
\label{N01eff}
\ea
The Neumann coefficients
$N_{01}$ are therefore related to $N_{00}$. The $N_{11}$'s are
also related to $N_{00}$:
\be
N_{11}^{12}=N_{11}^{13}=N_{11}^{23}
=\frac{a_1^2}{3}=\frac{1}{a^2}=e^{2N_{00}}.
\label{N11eff}
\ee

Using the redefined Neumann coefficient
(\ref{N00eff}), (\ref{N01eff}), and
(\ref{N11eff}) one finds that
the conditions (\ref{con1})--(\ref{con4}) are
satisfied. We thus conclude that any symmetric three string vertex
is compatible with the cubic interaction of the effective action.
\\

We end this section  showing that in fact any cyclic
three string vertex is compatible with the effective action.
In the following we show that the
absence of a $z^2$ term in $O(z)$
(\ref{a1}) (namely,  $a_2=0$) is necessary
only for the redundant
conditions (\ref{con1}) and  (\ref{con2}) to hold.
If $a_2\neq 0$, one finds that the only difference with respect to the
symmetric vertex is in $N_{01}^{rr}$
(before using momentum conservation):
\ba
N_{01}^{rr}&=&
-i\oint\frac{dz}{2\pi i}\frac{1}{z}\frac{h'_r(z)}{h_r(z)-h_r(0)}
\nonumber\\
&=&-i\oint\frac{dz}{2\pi i}\frac{1}{a_1 z^2}(a_1+(a_1^2+2a_2)z+...)
(1-(\frac{1}{2}a_1+\frac{a_2}{a_1})z+...) \nonumber\\
&=&-i(\frac{1}{2}a_1+\frac{a_2}{a_1}).
\label{00}
\ea
As before, $N_{01}^{rs}$ can be redefined using momentum
conservation s.t. $N_{01}^{rr}=0$. One finds
\ba
{\rm redefined}\;\; N_{01} &:& \nonumber\\
N_{01}^{21}=N_{01}^{32}=N_{01}^{13}
&=& \frac{1}{2}e^{N_{00}}+\frac{ia_2}{a_1},
\nonumber\\
N_{01}^{12}=N_{01}^{23}=N_{01}^{31}
&=&-\frac{1}{2}e^{N_{00}}+\frac{ia_2}{a_1},
\nonumber\\   N_{01}^{rr}=0.
\label{2/1}
\ea
(In appendix C it is shown that
for the SCSV vertex $a_1=i\sqrt{3}$,
$a_2=i\frac{\sqrt{3}}{2}$, and using
(\ref{2/1}) one recovers the Neumann coefficients given in section 3
(\ref{NCSV}).)
Finally, the  Neumann coefficients of a cyclic vertex given by
(\ref{N00eff}), (\ref{N11eff}) and (\ref{2/1}) satisfy conditions
(\ref{con3}) and (\ref{con4}),
and compatibility with the effective action
is  guaranteed.

\section{Effective Action and String Field Theory of the $N=2$ String}
\setcounter{equation}{0}
In this section we sketch the construction of the three $N=2$ closed
string vertex and
derive from it the effective action of  $N=2$ strings in toroidal
backgrounds. The following
discussing is far from complete as we ignore the ghost dependence and
picture changing. Yet, it is sufficient for comparison with the
effective action.

The critical $N=2$
string has perhaps the simplest vertex operator algebra of any
string theory.
This is because, unlike other string theories, it contains a
finite number of physical degrees of freedom~\cite{N=2,OV}.
In the
toroidal background $T^{2,2}$, physical operators are specified by the
lorentzian momenta $p$~\cite{GS}:
\be
p=(p_{Ls},p_{Rt};p_{Lt},p_{Rs})\in \Gamma^{4;4},
\label{p}
\ee
where $\Gamma^{4;4}$ is an even self--dual lorentzian lattice
with signature
$(4;4)$; $p_{Ii}$ is a complex vector, $I=L(R)$ for left--handed
(right--handed) momenta, and $i=s(t)$ for space--like (time--like)
components.

When both
$p_L\neq 0$ and $p_R\neq 0$ the only vertex operators that may become
physical at some point in the moduli space of toroidal backgrounds are
\be
V_p(X,\Xb)=e^{i(p\cdot \Xb + \pb\cdot X)},\qquad p^2=0,
\label{VpX}
\ee
where $p^2$ is defined by the lorentzian scalar product (\ref{pq}),
namely,
$p$ is a non--zero null vector of the $(4;4)$ lattice.
In (\ref{VpX}) the dot product is
\be
p\cdot \Xb \equiv p_L \Xb_L + p_R \Xb_R,
\ee
where $\pb$ is the complex conjugate of $p$, $p\Xb$ is the
lorentzian scalar product (\ref{pq}), and
$X_L^i$ and $X_R^i$ are the  $N=2$ (holomorphic and anti-holomorphic)
chiral superfields
\ba
X^i_L(Z,\t^-)&=&x^i_L(Z)+\psi^i_L(Z)\t^-,\qquad
X^i_R(\bar{Z},\tb^-)=x^i_R(\bar{Z})+\psi^i_R(\bar{Z})\tb^-, \nonumber\\
Z&=&z-\t^+ \t^-,  \qquad \bar{Z}=\bar{z}-\tb^+ \tb^- .
\ea
The superfield $\Xb$ is obtained from $X$ by complex conjugation
(in field space)
together with interchanging $\t^+\leftrightarrow \t^-$.

The upper component of the $N=2$ superfield $V_p(X,\Xb)$ (\ref{VpX}) is
\ba
V_p(z,\bar{z})&=&
(ip_L\partial\bar{x}_L-i\pb_L \partial x_L
-(p_L\bar{\psi}_L)(\pb_L\psi_L))
\nonumber\\
&\times&
(ip_R\partial\bar{x}_R-i\pb_R \partial x_R -(p_R\bar{\psi}_R)(\pb_R\psi_R))
e^{i(p\cdot \bar{x} +\pb\cdot x)}.
\label{Vpz}
\ea
The operator (\ref{Vpz}) is on--shell when both $p_L^2=0$ and $p_R^2=0$.
Given $p\in \Gamma^{4;4}$ there is a point in
the space of toroidal backgrounds
where $V_p$ is on--shell $iff$ $p^2=0$ (see ref. \cite{GS}),
and therefore, the operators (\ref{VpX}) are the set of
all operators  with
$p_L\neq 0$ and $p_R\neq 0$ that may become physical.

When $p=0$ there are discrete states \cite{B,GS}
\be
V_{ij}(z,\bar{z})=\frac{1}{2}(\partial x^i \bar{\partial} \bar{x}^j +
       \bar{\partial} x^i \partial \bar{x}^j).
\label{Vij}
\ee
The discrete states (\ref{Vij})
generate deformations of the metric moduli, and
are on--shell  at any background. They can be formed from off--shell
$V_p$'s by choosing an appropriate normalization as $p$ goes to
zero~\cite{GS}.

In addition to $V_{ij}$, there are discrete states that appear in
particular toroidal backgrounds~\cite{GS}. They can be formed
from off--shell
$V_p$'s by choosing an appropriate normalization as $p_R\rightarrow 0$
while $p_L^2=0, p_L\neq 0$ (and similarly for $p_L\rightarrow 0$
while $p_R^2=0, p_R\neq 0$). They are given by the vertex operators
\ba
J^i_{(p_L,0)}&=&
(ip_L\partial\bar{x}_L-i\pb_L \partial x_L-(p_L\bar{\psi}_L)(\pb_L\psi_L))
\;\bar{\partial} x_R^i \;  e^{i(p_L\bar{x}_L +\pb_L x_L)},
\nonumber \\
\bar{J}^i_{(0,p_R)}&=&
\partial x_L^i
(ip_R\partial\bar{x}_R-i\pb_R \partial x_R -(p_R\bar{\psi}_R)(\pb_R\psi_R))
e^{i(p_R\bar{x}_R +\pb_R x_R)}.
\label{J}
\ea
The discrete states in (\ref{J}) give rise to exact gauge symmetries.

We are now ready to expand the $N=2$ string field in terms of
all the Fock space
states that may become  physical modes in some toroidal background. In
the 0--picture:
\ba
|\Phi>&=&\sum_{p^2=0} \phi_p V_p |0> + \sum_{ij} G^{ij}V_{ij}|0>
\nonumber \\
&+& \sum_{i,p^2=0} A_{ip} J^i_{p}|0> +
    \sum_{i,p^2=0} \bar{A}_{ip} \bar{J}^i_{p}|0>.
\label{Phi}
\ea
Here $J^i_p$ and $ \bar{J}^i_{p}$ are the extensions of (\ref{J}) to
off--shell momenta $p$.
The terms involving the gauge fields $A_{ip}$ ($\bar{A}_{ip}$) are summed
over all momenta $p\in \Gamma^{4;4}$ that may become of the form
$(p_L,0),\;p_L^2=0, p_L\neq 0$ ($(0,p_R),\; p_R^2=0, p_R\neq 0$)
at some point of the moduli space of toroidal backgrounds; those are also
the set of non--zero null vectors of $\Gamma^{4;4}$, and therefore we sum
over $p^2=0$ in (\ref{Phi}).\footnote{
We can extend the sums in (\ref{Phi}) to all $p\in \Gamma^{4;4}$; the modes
with $p^2\neq 0$ will then be eliminated by the $\delta(L_0-\bar{L}_0,0)$
in the three string vertex. With this extension we have a one to one
correspondence between modes of the string field and elements of the
volume--preserving diffeomorphism algebra of $T^{4;4}$ (see
ref.~\cite{GS}).
}

Following the discussion in section 3,
the three $N=2$ string vertex is determined by the parameter
$a$ in (\ref{az})
(because higher oscillations of the $N=2$ string are never physical),
and the cubic interaction for a cyclic vertex is
\ba
S_3(N=2)&=&S_{\phi\phi\phi}+S_{G\phi\phi}+S_{A\phi\phi},\nonumber\\
S_{\phi\phi\phi}&=&\frac{1}{3}\sum_{p_1,p_2,p_3}
a^{-\frac{1}{2}\sum_{r=1}^3 p_{Lr}^2}
\bar{a}^{-\frac{1}{2}\sum_{r=1}^3 p_{Rr}^2}
\; \delta_{p_1+p_2+p_3,0}\nonumber\\ &\times&
\frac{1}{4}(p_{L2}\pb_{L3}-p_{L3}\pb_{L2}) (p_{R2}\pb_{R3}-p_{R3}\pb_{R2})
\phi_{p_1} \phi_{p_2} \phi_{p_3},
\nonumber \\ S_{G\phi\phi}&=& \frac{1}{4}
\sum_{p} a^{-p_L^2} \bar{a}^{-p_R^2}
(p_L^i \pb_R^j + \pb_L^i p_R^j)G_{ij} \phi_p \phi_{-p},
\nonumber\\
S_{A\phi\phi}&=& \frac{1}{3}\sum_{p_1,p_2,p_3}
a^{-\frac{1}{2}\sum_{r=1}^3 p_{Lr}^2}
\bar{a}^{-\frac{1}{2}\sum_{r=1}^3 p_{Rr}^2} \;
\delta_{p_1+p_2+p_3,0} \nonumber\\
&\times & \frac{1}{2}
{[}(p_{L2}\pb_{L3}-p_{L3}\pb_{L2}) p_{R1}^i A_{ip_1}\phi_{p_2}\phi_{p_3}
\nonumber\\ &+&
(p_{R2}\pb_{R3}-p_{R3}\pb_{R2}) p_{L1}^i
\bar{A}_{ip_1}\phi_{p_2}\phi_{p_3}{]}.
\label{S3N2}
\ea
The sums in (\ref{S3N2}) are
only over the momenta appropriate to each of the
fields $\phi_p, A_{ip}, \bar{A}_{ip}$ as discussed above.
The first term in (\ref{S3N2}) arises from the off--shell three point
function~\footnote
{
More precisely, one calculates tree--level correlation functions by
inserting $V_p$ at $n$ points on the sphere, and integrating over their
positions modulo global superconformal transformations.
One then finds that
two of the $V_p$'s in (\ref{p1p2p3}) are
in the $(-1,-1)$--picture, namely,
$V_p^{(-1,-1)}\equiv c\bar{c}e^{-\phi_+}e^{-\phi_-}e^{-\bar{\phi}_+}
e^{-\bar{\phi}_-} e^{i(p\cdot\bar{x}+\pb\cdot x)}$, where $c$ is the
spin $-1$ ghost of the $(b,c)$ system, and $\phi_+,\phi_-$ are the two
scalars used in bosonizing the spin $3/2,-1/2$
ghosts $(\beta^{\pm},\gamma^{\pm})$
of the $N=2$ string \cite{Li}.
}
(\ref{Vijk})
\ba
V_{p_1,p_2,p_3}&=&<0|h_3(V_{p_1}(0))h_2(V_{p_2}(0))h_1(V_{p_1}(0))|0>
\nonumber\\
&=& \frac{1}{4} a^{-\frac{1}{2}\sum_{r=1}^3 p_{Lr}^2}
\bar{a}^{-\frac{1}{2}\sum_{r=1}^3 p_{Rr}^2}\nonumber\\
&\times& \delta_{p_1+p_2+p_3,0}
(p_{L2}\pb_{L3}-p_{L3}\pb_{L2}) (p_{R2}\pb_{R3}-p_{R3}\pb_{R2}).
\label{p1p2p3}
\ea
The second term in (\ref{S3N2}) arises from the off--shell three point
function
\ba
V_{(ij),p,-p}&=&<0|h_3(V_{ij}(0))h_2(V_{p}(0))h_1(V_{-p}(0))|0>
\nonumber\\
&=& \frac{1}{2} a^{-p_L^2} \bar{a}^{-p_R^2}
(p_L^i \pb_R^j + \pb_L^i p_R^j).
\ea
The third term in (\ref{S3N2}) arises from the off--shell three point
functions
\ba
V_{(ip_1),p_2,p_3}
&=&<0|h_3(J_{ip_1}(0))h_2(V_{p_2}(0))h_1(V_{p_3}(0))|0>
\nonumber\\
&=&\frac{1}{2} a^{-\frac{1}{2}\sum_{r=1}^3 p_{Lr}^2}
\bar{a}^{-\frac{1}{2}\sum_{r=1}^3 p_{Rr}^2}
\; \delta_{p_1+p_2+p_3,0}
(p_{L2}\pb_{L3}-p_{L3}\pb_{L2}) p_{R1}^i,\nonumber\\
V_{(\bar{i}p_1),p_2,p_3}
&=&<0|h_3(\bar{J}_{ip_1}(0))h_2(V_{p_2}(0))h_1(V_{p_3}(0))|0>
\nonumber\\
&=&\frac{1}{2} a^{-\frac{1}{2}\sum_{r=1}^3 p_{Lr}^2}
\bar{a}^{-\frac{1}{2}\sum_{r=1}^3 p_{Rr}^2}
\; \delta_{p_1+p_2+p_3,0}\; p_{L1}^i (p_{R2}\pb_{R3}-p_{R3}\pb_{R2}).
\nonumber\\ {}
\ea

At the decompactification limit the target--space is $R^{2,2}$, and
$p_L^i=p_R^i\equiv p^i$ is a continuous space--time
momentum. After performing a Fourier transform from momentum space to
coordinate space $S_3(N=2)$ becomes
\ba
S_3^{\rm dec}&=&\int d^2 x_1 d^2 x_2 \; \frac{1}{2}
G^{i\bar{j}}
(e^{-log|a|\eta^{k\bar{l}}\partial_k\partial_{\bar{l}}}\partial_i\phi)
(e^{-log|a|\eta^{k\bar{l}}
\partial_k\partial_{\bar{l}}}\partial_{\bar{j}}\phi)
\nonumber \\
&+&\frac{1}{3}\e^{ij}\e^{\bar{i}\bar{j}}
(e^{-log|a|\eta^{k\bar{l}}\partial_k\partial_{\bar{l}}}\phi)
(e^{-log|a|\eta^{k\bar{l}}\partial_k\partial_{\bar{l}}}
\partial_j\partial_{\bar{i}}\phi)
(e^{-log|a|\eta^{k\bar{l}}\partial_k\partial_{\bar{l}}}
\partial_i\partial_{\bar{j}}\phi)
\nonumber\\
&+&\frac{1}{3}\eta^{i\bar{j}}\eta^{\bar{i}j}
(e^{-log|a|\eta^{k\bar{l}}\partial_k\partial_{\bar{l}}}\phi)
(e^{-log|a|\eta^{k\bar{l}}\partial_k\partial_{\bar{l}}}
\partial_i\partial_{\bar{j}}\phi)
(e^{-log|a|\eta^{k\bar{l}}\partial_k\partial_{\bar{l}}}
\partial_j\partial_{\bar{i}}\phi).\nonumber\\{}
\label{Sdec}
\ea

The factors $e^{-log|a|\eta^{k\bar{l}}\partial_k\partial_{\bar{l}}}$
in (\ref{Sdec}) can be eliminated by an
appropriate field redefinition (see
explanation in section 3)~\footnote
{ The field redefinition (\ref{def}) is singular in the
decompactification  case due to the factor $\frac{1}{p^2}$,
but an appropriate regular field redefinition can be done by replacing
(\ref{def}) with
$ Z_p=z_p-\frac{g}{3}\sum_{q}
\left( \frac{e^{-Ap^2}-1}{p^2}\right) z_q z_{-p-q}
{[}f(p,q) e^{-A(q^2+(p+q)^2)}+f(q,p)e^{-A(p+q)^2}+f(-p-q,q){]}
+o(z^3)$. The factor $\frac{e^{-Ap^2}-1}{p^2}$ is not singular in such a
field redefinition.
}.
Such a field redefinition introduces higher
order terms in the new field $\phi$.
After eliminating the exponents, and
for $G^{i\bar{j}}=\eta^{i\bar{j}}$, one can get rid of the third term in
(\ref{Sdec}) by an additional field  redefinition
\be
\phi\rightarrow
\phi+\frac{1}{3}\eta^{i\bar{j}}\phi\partial_i\partial_{\bar{j}}\phi.
\label{pred}
\ee
The field redefinition (\ref{pred}) also introduces higher order terms in
the new field $\phi$.
Now, dropping four and higher order terms in
(\ref{Sdec}) one recovers the effective action presented in \cite{OV}:
\be
S_{\rm eff}=\int d^2 x_1 d^2 x_2 \left( \frac{1}{2}
\eta^{i\bar{j}} \partial_i\phi\partial_{\bar{j}}\phi
+\frac{1}{3}\e^{ij}\e^{\bar{i}\bar{j}}\phi
\partial_j\partial_{\bar{i}}\phi
\partial_i\partial_{\bar{j}}\phi \right).
\label{2eff}
\ee

The gauge fields do not appear explicitly in (\ref{Sdec}),(\ref{2eff});
they can be
introduced via a K\"ahler transformation~\cite{GS}
\be
\partial\phi\rightarrow\partial\phi+A, \qquad
\bar{\partial}\phi\rightarrow\bar{\partial}\phi+\bar{A}.
\label{kahler}
\ee
In Minkowski background the transformation (\ref{kahler}) introduces
total derivatives.

So far, we have discussed only the three--point functions.
It turns out that four--point and probably also higher--point amplitudes of
the operators $V_p$ (\ref{VpX}) vanish
\cite{OV,BGI,Li}.
This fact is already
predicted by the  effective action (\ref{2eff}) and it is thus
believed to be correct to all orders.
However, in the action
$S_3^{\rm dec}$ (\ref{Sdec}) we get extra terms of the form
$\eta ... \eta \phi...\phi$.
These terms should be canceled against higher order
interactions of (off--shell) $V_p$'s and discrete states $V_{ij}$,
when including the
higher string vertices of the $N=2$ closed  string field theory.

Before ending this section, we remark that the kinetic term in the
effective action (\ref{2eff}) can be derived, alternatively, from
the quadratic term of the $N=2$ string field theory. Schematically,
\be
{}_{12}<R|Q|\Phi>_1|\Phi>_2 \;\sim \;
\sum_p (p_L\pb_L+p_R\pb_R)\phi_p\phi_{-p},
\ee
where $|R>_{12}$ is the two string reflector, and  $Q$ is the
$N=2$ BRST operator.

\section{Summary and Discussion}
\setcounter{equation}{0}
In this section we summarize the main points and present a
discussion about this work.
The structure of the scalar potential in gauged
$D=4$, $N=4$ supergravity coupled
to matter is fixed once the gauge algebra is given.
This was used in order
to derive a non--trivial check for the compatibility of SFT and effective
actions to cubic order. It would be interesting to check the compatibility
of higher order interactions as well. For that purpose one needs to work
with the four (and higher) closed string vertex,
and integrate out massive modes to derive an
effective action.
Assuming that the effective action is compatible with SFT to all orders we
hope to use $D=4$ supergravity to study some important problems in SFT.
This is left for a future work; here we discuss some of the open problems.

By inserting the quadratic constraint (\ref{quacon}) into
the scalar potential (\ref{V}), $V(Z)$  becomes non--polynomial in the
physical fields. The structure of  $V(Z)$ might be used to understand the
non--polynomiality of closed string field theory. Moreover, the effective
action of $N=4$ heterotic strings is completely duality invariant when the
gauge algebra is the lorentzian lattice algebra of the Narain lattice
$\Gamma^{6,22}$ \cite{GP}. The infinite dimensional gauge symmetry is
spontaneously broken at any point in the moduli space of toroidal
backgrounds. The generalized duality transformations are residual discrete
symmetries of the broken
gauge algebra \cite{GMR}.
As expected, a similar phenomenon occurs in string field theory:
target-space duality as a symmetry of string field theory was discussed
recently for
the $\a=p^+$ HIKKO string field theory in toroidal backgrounds in ref.
\cite{KZ}.

The  nature of the infinite gauge algebra
suggests that (for toroidal backgrounds)
the underlying invariance principle of SFT has to do with an algebra
defined on the Narain torus of compactification.
Such is the case also for the $N=2$ strings, for which an underlying
off--shell
algebra  isomorphic to vdiff($T^{4,4}$) was suggested \cite{GS}.

Another important point is the question of background independence. The
effective action of  $N=4$ heterotic strings is manifestly background
independent, namely, it is independent of the toroidal background around
which it is formulated. Different backgrounds correspond to different
classical solutions, and are interpolated by changing
expectation values of  scalar fields at the minimum of the potential.
The infinite gauge algebras in different backgrounds are isomorphic,
although
there are different unbroken gauge groups in different
backgrounds. We expect a
similar phenomenon to occur in string field theory.
However, in SFT
the spin one ultra--massive gauge fields are replaced with the
higher spin stringy modes, and revealing the underlying symmetry is
harder.\\

We have shown that quadratic and cubic interactions in SFT of the
fields $Z_p^a$ and $Z_{p=0}^{aI}$ are
compatible with the $N=4$ supergravity effective
action for any cyclic vertex.
Including the fields $Z_p^{aI},\; p\neq 0$
in the cubic interaction derived from SFT
(\ref{S(Z)}) (in addition to $p=0$)  gives some interesting
results. For example, it gives the leading order of a $D=10$
supergravity in the decompactification limit,
as was speculated in~\cite{GP}.
It would be interesting to study further the
structure of cubic interactions that involve these  fields.

A cyclic vertex is
defined by a  single map $h(z)=e^{O(z)}$, where
$O(z)=a_1 z +a_2 z^2+o(z^3)$ (\ref{a1}).
The parameter $a_1$ in the three string vertex
gives rise to a scale factor
$e^{-{\rm log}(\frac{\sqrt{3}}{|a_1|})\sum_{r=1}^3 p_{L(r)}^2}$ (\ref{eE})
in the cubic interaction.
Although apparently  this factor is not compatible with
effective actions, there is a (non--linear) field redefinition (\ref{def})
which gets rid of it to cubic order, and therefore, a
cyclic three string vertex is compatible with effective
actions for any scale factor.
It would be interesting to study compatibility of the quadratic and cubic
interactions
using a general three string vertex.

The parameter $a_1$ specify the scaling factor of the map from
the three unit discs (or semi--infinite cylinders) to the sphere.
An overlap Witten vertex (namely,
the three discs cover precisely the complete sphere) has a scale factor
$a_1=4/3$ (see appendix B (\ref{Ba})).
There are good reasons to believe that an overlap vertex is preferable,
but we did not find a restriction like that arising from cubic order
effective actions. \\

For the $N=2$ string we have
included in the  string field $|\Phi>$ (\ref{Phi})
all the modes that may become physical
somewhere in the moduli space of toroidal backgrounds. Therefore, assuming
this is all one needs for a consistent string field theory,
the $N=2$ string field theory
is equivalent to its effective action.
We have found that the effective action in Minkowski
background agrees with the
covariant closed string field theory to cubic order if we make an
appropriate field redefinition. The non--linear
field redefinition introduces higher order interactions.
The higher order interactions found by studying the three
string vertex must be canceled  when including the higher string vertices
of the (non--polynomial) covariant string field theory.

Finally,
it is important to understand better the role of the discrete states in
$N=2$ string field theory, and the way they fit into a background
independent formulation. A similar difficulty arises when studying
effective actions of the $d=2$ string \cite{WKPWZ}.
It may also be interesting to generalize this work to curved backgrounds,
both  for the $N=2$ strings and the $N=1$ heterotic string.\\

\vskip .05in \noindent
Acknowledgements \vskip .05in \noindent
I would like to thank O. Lechtenfeld, A. Shapere, and E. Witten for
discussions, and especially B. Zwiebach for a lot of help and for a
critical reading of the manuscript.
This work is supported in part by DOE grant no.
DE--FG02--90ER40542.

\section*{Appendix A - Useful Formulas}
\renewcommand{\theequation}{A.\arabic{equation}}
\setcounter{equation}{0}
In this appendix we present some useful formulas that help in particular
in deriving eq. (\ref{ZZZV}).

Hermitian conjugation of the oscillators $\a$ and $\bar{\a}$ in (\ref{E})
is given by
\be
\a^{\dagger}_n=\a_{-n},\qquad
\bar{\a}^{\dagger}_n=\bar{\a}_{-n}.
\ee
The commutation relations of the oscillators with $E_{123}^{\dagger}$
(the hermitian conjugate of (\ref{E})) are:
\ba
[\a_n^{a(r)},E_{123}^{\dagger}]
&=&\sum_{s}n(\sum_{m>0}N_{nm}^{rs}\a_{-m}^{a(s)}
                        +N_{n0}^{rs}\a_0^{a(s)}),\nonumber\\
{[}\bar\a_n^{I(r)},E_{123}^{\dagger}{]}&=&
\sum_{s}n(\sum_{m>0}\bar{N}_{nm}^{rs}\bar{\a}_{-m}^{I(s)}
                         +\bar{N}_{n0}^{rs}\bar{\a}_0^{I(s)}),
\label{[aE]}
\ea
where $n>0$ in (\ref{[aE]}). Also
\ba
{[}\a_n^{a(r)},e^{E_{123}^{\dagger}}{]}
&=&{[}\a_n^{a(r)},E_{123}^{\dagger}{]}e^{E_{123}^{\dagger}},
\nonumber\\
{[}\bar{\a}_n^{I(r)},e^{E_{123}^{\dagger}}{]}
&=&{[}\bar{\a}_n^{I(r)},E_{123}^{\dagger}{]}
e^{E_{123}^{\dagger}},
\label{[a,eE]}
\ea
where $n>0$ in (\ref{[a,eE]}).

The zero modes $\a_0$ and $\bar{\a}_0$ are operators which pick up values
in the Narain lattice of compactification in the following way:
\ba
{}_{1}<p|\;{}_{2}<p|\;{}_{3}<p|\a_0^{a(r)}|0>_{123}=
\delta_{p_1+p_2+p_3,0}\; p_{L(r)}^a,
\nonumber\\
{}_{1}<p|\;{}_{2}<p|\;{}_{3}<p|\bar{\a}_0^{I(r)}|0>_{123}=
\delta_{p_1+p_2+p_3,0}\; p_{R(r)}^I.
\label{pa0}
\ea
(In the literature $\a_0$ is sometimes identified with $ip$,
and this gives
rise to relative $i$ factors in the formulas used in different works.)

We end this appendix with a simple symmetry of  the Neumann coefficients,
which is used repeatedly in the calculations:
\be
N_{nm}^{rs}=N_{mn}^{sr}.
\ee

\section*{Appendix B - Witten Vertex}
\renewcommand{\theequation}{B.\arabic{equation}}
\setcounter{equation}{0}
In this appendix we derive the map $h(z)$ in eq. (\ref{h123}) for Witten
vertex and compute the Neumann coefficients.

The three string Witten vertex is a symmetric overlap vertex,
namely, the images of the three unit discs mapped to the sphere by
$h_1$, $h_2$ and $h_3$ in (\ref{h123})
cover precisely the complete sphere.
With the three punctures at $z=1,\r,\rb$,
where $\r=e^{\frac{2\pi i}{3}}$,
the map $h(z)$ is manifestly cyclic symmetric,
and thus map the unit disc $|z|\leq 1$  onto one third of the complex plan
containing the puncture at $z=1$ (see figure 1), namely
\be
h(z)=\left(\frac{1+z}{1-z}\right)^{\frac{2}{3}}.
\label{Bh}
\ee
Expanding in $z$
\be
h(z)=1+\frac{4}{3}z+\frac{8}{9}z^2+o(z^3),
\ee
and comparing with the parametrization
$h(z)=1+a_1 z+(\frac{1}{2}a_1^2+a_2)z^2+o(z^3)$ in (\ref{a1}) one finds
\be
a_1=\frac{4}{3}, \qquad a_2=0.
\label{Ba}
\ee
{}From eq. (\ref{witten}) one finds $a=\frac{3\sqrt{3}}{4}$, which is
compatible with the relation (\ref{aa1}).

Inserting (\ref{Ba}) in (\ref{N00eff}),(\ref{N11eff}) and (\ref{2/1})
one finds the Neumann coefficients presented in section 3 (\ref{n00}),
(\ref{n01}) and (\ref{n11}).

\section*{Appendix C - SCSV Vertex}
\renewcommand{\theequation}{C.\arabic{equation}}
\setcounter{equation}{0}
In this appendix we derive the map $h(z)$ in eq. (\ref{h123}) for SCSV
vertex and compute the Neumann coefficients.

The SCSV vertex is a (non--symmetric) cyclic vertex defined by the map
$h(z)=z$ when the three punctures are at $z=0,1,\infty$. To derive the
SCSV vertex with the three punctures at $z=1,\r,\rb$, where
$\r=e^{\frac{2\pi i}{3}}$, we act on $h(z)$ with the $SL(2,C)$
transformation which maps $0\rightarrow 1$, $1\rightarrow \r$ and
$\infty \rightarrow \rb$. One finds
\be
h(z)=\frac{1+\r z}{1+\rb z},
\label{Ch}
\ee
and expanding in $z$
\be
h(z)=1+i\sqrt{3}z+(i\frac{\sqrt{3}}{2}-\frac{3}{2})z^2+o(z^3).
\ee
Comparing with the parametrization
$h(z)=1+a_1 z+(\frac{1}{2}a_1^2+a_2)z^2+o(z^3)$ in (\ref{a1}) one finds
\be
a_1=i\sqrt{3}, \qquad a_2=i\frac{\sqrt{3}}{2}.
\label{Ca}
\ee
Inserting (\ref{Ca}) in (\ref{N00eff}),(\ref{N11eff}) and (\ref{2/1})
one finds the Neumann coefficients
\ba
N_{11}^{rs}&=&\frac{a_1^2}{3}=-1,\qquad {\rm for}\;\; r\not= s
\nonumber \\
N_{01}^{21}&=&N_{01}^{32}=N_{01}^{13}=
\frac{a_1}{2\sqrt{3}}+\frac{ia_2}{a_1}=i, \nonumber \\
N_{01}^{21}&=&N_{01}^{23}=N_{01}^{31}=
-\frac{a_1}{2\sqrt{3}}+\frac{ia_2}{a_1}=0, \nonumber\\
N_{00}^{rs}&=&\delta^{rs} {\rm log}\frac{a_1}{\sqrt{3}}
=\delta^{rs}{\rm log}i,
\label{CN}
\ea
and the rest of the Neumann coefficients are $0$. The $N's$ in (\ref{CN})
are the Neumann coefficients presented in section 3 (\ref{NCSV}).

\newpage

\noindent
{\bf FIGURE CAPTIONS}\\

\noindent
{\bf Figure 1}:\\
The maps $h_1,h_2,h_3$ from the coordinate systems $z_1,z_2,z_3$
to the $z$--plane for Witten vertex with punctures at $z=1,\r,\rb$,
$\r=e^{\frac{2\pi i}{3}}$.
\end{document}